\documentstyle[11pt, amssymb]{article}                                

\def\AFOUR{%
\setlength{\textheight}{9.0in}%
\setlength{\textwidth}{5.75in}%
\setlength{\topmargin}{-0.375in}%
\hoffset=-.5in%
\renewcommand{\baselinestretch}{1.17}%
\setlength{\parskip}{6pt plus 2pt}%
}

\AFOUR

\makeatletter
\def\section{\@startsection {section}{1}{\z@}{-3.5ex plus -1ex minus
 -.2ex}{2.3ex plus .2ex}{\large\bf}}
\def\subsection{\@startsection{subsection}{2}{\z@}{-3.25ex plus -1ex
minus
 -.2ex}{1.5ex plus .2ex}{\normalsize\bf}}
\makeatother

\makeatletter
\@addtoreset{equation}{section}

\makeatother

\def\overleftrightarrow#1{\vbox{\ialign{##\crcr
\leftrightarrow\crcr\noalign{\kern-1pt\nointerlineskip}
\hfil\displaystyle{#1}\hfil\crcr}}}

\newcommand{\nc}{\newcommand}
\newcommand{\rnc}{\renewcommand}

\nc{\be}{\begin{equation}}
\nc{\ee}{\end{equation}}
\nc{\bea}{\begin{eqnarray}}
\nc{\eea}{\end{eqnarray}}

\rnc{\a}{\alpha}
\rnc{\b}{\beta}
\nc{\adot}{\dot{\alpha}}
\nc{\bdot}{\dot{\beta}}
\rnc{\L}{\Lambda}
\rnc{\l}{\lambda}
\nc{\CA}{\mathcal{A}}
\nc{\CB}{\mathcal{B}}
\nc{\CM}{\mathcal{M}}
\nc{\CN}{\mathcal{N}}
\nc{\CK}{\mathcal{K}}
\nc{\BAB}{B^{\mathcal{A}\mathcal{B}}}
\nc{\ab}{\bar{\a}}
\nc{\lb}{\bar{\l}}
\nc{\etab}{\bar{\eta}}

\nc{\tr}{\mathop{\mbox{tr}}\nolimits}

\begin{document}
\global\parskip=4pt


\begin{titlepage}

\begin{flushright}
\end{flushright}
\vspace*{0.5in}
\begin{center}
{\LARGE{\sc
Instantons in ${\CN}=2$ $Sp(N)$ Superconformal Gauge Theories and the
  AdS/CFT Correspondence }}\\
\vskip .3in
{\large\sc E. Gava}\\ 
{\it INFN, ICTP and SISSA, Trieste, Italy}\\
\vspace{.2in}
{\sc K.S.\ Narain}\\
{\it ICTP, Strada Costiera 11, 34014 Trieste, Italy}\\
\vspace{.2in}
{\large\sc M.H. Sarmadi}\\
{\it Institute for Studies in Theoretical Physics and Mathematics IPM,\\ 
 P.O. Box 19395-5531, Tehran, Iran}\\

\end{center}

\begin{abstract}
\noindent 
We study, using ADHM construction, instanton effects
in an ${\CN}=2$ superconformal $Sp(N)$ gauge theory,
arising as effective field theory on a system 
of $N$ D-3-branes near an orientifold 7-plane and 8
D-7-branes in type I' string theory. We work out
the measure for the collective coordinates of multi-instantons 
in the gauge theory and compare
with the measure for the collective coordinates of 
$(-1)$-branes in the presence of 3- and 7-branes
in type I' theory. We analyse the large-$N$
limit of the measure and find that it admits two classes of saddle 
points: In the first class the space of collective coordinates has
the geometry of $AdS_5\times S^3$ which on the string theory side 
has the interpretation of the D-instantons being stuck on the 7-branes
and therefore the  resulting moduli space being $AdS_5\times S^3$,
In the second class the geometry is $AdS_5\times S^5/Z_2$ and on the
string theory side it means that the D-instantons  
are free to move in the 10-dimensional bulk. 
We discuss in detail a correlator of four $O(8)$ flavour currents
on the Yang-Mills side, which receives contributions from
the first type of saddle points only, and show that it matches with 
the correlator obtained from $F^4$ coupling on the string theory
side, which receives contribution from D-instantons, in perfect 
accord with the AdS/CFT correspondence. In particular we observe 
that the sectors with odd number of instantons give contribution
to an $O(8)$-odd invariant coupling, thereby breaking $O(8)$ down
to $SO(8)$ in type I' string theory.  We finally discuss correlators 
related to $R^4$ , which receive contributions from both saddle points.

\end{abstract}
\end{titlepage}
\makeatother


\setcounter{footnote}{0}

\section{Introduction}

Dualities, and in particular strong-weak coupling dualities, have
shed light on our understanding of the dynamics of string
theories and Yang-mills theories. More recently the conjecture of
Maldacena \cite{Mal} relates the physics of certain conformally invariant large 
$N$ 
Yang-Mills
theories living on 3-branes in various string theories to that of
the corresponding bulk string theories in the near horizon geometry
(AdS geometry) of the 3-branes. Many detailed checks
have so far confirmed this conjecture and have at the same time
provided new insights into both the physics of Yang-Mills theories
as well as that of string theories. In a beautiful paper 
 \cite{Dorey}, in the context of ${\CN}=4$ Yang-Mills theories, 
 this equivalence was extended to $e^{-N}$ orders. On the
Yang-Mills side they appear as instanton effects, while in the
string theory they correspond to certain higher derivative couplings
induced by stringy instantons. What was remarkable was that in the large
$N$ limit, the moduli space of $SU(N)$ multi-instantons collapses into 
the "center of mass" moduli that live on $AdS_5 \times S^5$ as predicted by 
AdS/ CFT correspondence, and a set of "relative" moduli whose action
is that of a multi D-instanton action in the underlying IIB string
theory. It was known through the works \cite{gg, Banks} and \cite{Bianchi}
that the latter contributes to certain $R^4$ terms in the IIB theory (R
being the Riemann tensor). Also, it was shown that certain
correlators in the Yang-Mills instanton background, had a direct
correspondence with certain couplings appearing in the string theory
via the dictionary of the Yang-Mills operators and the bulk 
operators as given in \cite{Witten, Gubser} .

The purpose of this paper is to extend the results of ref. \cite{Dorey} to
the case of ${\CN}=2~ Sp(N)$ Yang-Mills theories that appear as the 
3-brane world volume
theory in the type I' model where the 3-branes are 
living at an orientifold 7-plane
together with 8 D-7-branes. The near horizon geometry of this system is that of 
$AdS_5 \times X_5$ where $X_5$ is a particular 
$Z_2$ modding of $S^5$. The 7-brane
world volume intersects with $X_5$ on an $S^3$. This model was first studied in
refs. \cite{Fayyaz1, Fayyaz2} following the earlier 
works \cite{Sen, Seiberg}. Subsequently the order $N$ 
corrections to the AdS/CFT trace anomaly in 
this model were analyzed in refs. \cite{Aharony,
Narain}.
In the bulk theory this correction was traced, in ref. \cite{Narain}, 
to the presence of an $R^2$ term in the 7-brane world
volume action. 

In the present work we will consider instantons in the 3-brane world
volume theory and show that, in the large $N$ limit, two types of saddle points
are relevant: Those for which  multi-instanton moduli space is $AdS_5 \times 
S^3$, and the 
fluctuations are given by the D-instanton action in type I' theory,
and those for which the multi-instanton moduli space is 
$AdS_5 \times S^5/Z_2$, and the 
fluctuations are given by the D-instanton action in type IIB theory. For
the former saddle points the interpretation on the string side is that
the D-instatons sit at a point on the the 7-branes whereas for the
latter ones they can be at a point in the whole 10 dimensional bulk.   
We will study certain correlation functions, in the
Yang-Mills theory, of the $O(8)$ flavour 
currents that couple to $O(8)$ gauge fields 
in the bulk via AdS/ CFT correspondence. 
We will see that in the large $N$ limit the first type of saddle points are 
relevant
for these correlation functions. Moreover, we will show that
these correlation functions  describe the bulk to boundary
propagators that connect $F^4$
vertex to the $O(8)$ currents on the boundary. 
The result then is obtained by the D-instanton
contributions to $F^4$ in type I' theory or equivalently D-string instanton
contributions in type I theory \cite{bachas1, gmnt} or world sheet
instantons at
string one-loop level in heterotic theory \cite{bachas,Lerche}. We will see
that both odd and even instanton numbers contribute to this correlator. 
We will also see that certain
correlators can receive contributions from instanton configurations whose
moduli space is the full 10-dimensional bulk spacetime $AdS_5\times S^5/Z_2$.

Some of the issues developed here have been recently discussed in
\cite{gut,bianchi1}.
Also instanton effects in $SU(N)$ superconformal ${\CN}=2$ gauge theories
have been discussed in \cite{hkm} and, in the context of orbifold AdS/CFT
correspondence, in \cite{hk}

The paper is organized as follows. In section 2 we review the 3-brane world 
volume
theory and describe the relevant ADHM construction 
of the multi-instantons. In section
3 we describe the ADHM measure in terms of 
integrals over the fields on $(-1)$-branes
in a system of $(-1)$-, 3- and 7-branes. We then go on in section 4
to find the saddle point
solutions in the large-$N$ limit and show that the integral collapses to a
"center of
mass" integral on $AdS_5 \times S^3$ (i.e. in the 7-brane world volume where
$O(8)$ gauge fields live) or $AdS_5\times S^5/Z_2$, together with integrals
over the fluctuations describing the relative
positions of D-instantons in type I' or  IIB theory
respectively. For the first saddle point, we will see that both odd and even
instantons contribute to the correlation 
function involving four $O(8)$ currents and give rise to odd and even 
quartic invariants of $O(8)$. In section 5 we show 
that the four $O(8)$ current correlators describe bulk to 
boundary propagators connecting 
$F^4$ vertex in $AdS_5$ to four $O(8)$ currents on the boundary of $AdS_5$. We 
will
explicitly show the emergence of the usual $t_8$ tensor in $F^4$ vertex. 
In section 6, we describe the heterotic string
computation of $F^4$ term for the odd invariant case since this has not 
appeared in the literature so far. We will also discuss $R^4$
couplings, which receive contributions from both saddle
points. Finally in
section 7 we make concluding remarks.

\section{ADHM construction of ${\CN}=2~ Sp(N)$ instantons}

The theory on the type I' 3-branes that we are considering here is
${\CN}=2~ Sp(N)$ Yang Mills theory with one hypermultiplet transforming
under the antisymmetric representation of $Sp(N)$ and a fundamental
hyper multiplet transforming as $(N,8)$ under $Sp(N)\times
O(8)$ with $O(8)$ being the flavour symmetry arising from the
7-branes. The following table includes the fields on the 3-brane
world volume together with their $SO(4)_I \times SO(4)_E \times
SO(2)$ transformation properties.  Here the subscript $I$ refers to
the $SO(4)$ of the world volume (writing $SO(4)=SU(2)^2$, we shall
label the quantum numbers by $\alpha, \dot{\alpha}$), $E$ refers to
the external $SO(4)$ which is part of the 7-brane world volume ($A$
and $Y$ label the analogous quantum numbers) and finally $SO(2)$ acts
on the space transverse to the 7-branes. 

\begin{center}
D$3$-brane  World-volume content
\end{center}
\begin{tabular}{lccccc}
Fields           &$SO(4)_I$ &$SO(4)_E$   &$SO(2)$      &$Sp(N)$   &$O(8)$ \\
$v_{\mu}$          &$(2,2)$   &$(1,1)$   &$0  $        &$N(2N+1)$ &$1$    \\
$\varphi^{\pm}$    &$(1,1)$   &$(1,1)$   &$\pm 2 $     &$N(2N+1)$ &$1$    \\
$\varphi^{AY}$     &$(1,1)$   &$(2,2)$   &$0  $        &$N(2N-1)$ &$1$    \\
$\l^A_\a$          &$(2,1)$   &$(2,1)$   &$+1 $        &$N(2N+1)$ &$1$    \\
$\lb^A_{\adot}$    &$(1,2)$   &$(2,1)$   &$-1 $        &$N(2N+1)$ &$1$    \\
$\l^Y_\a$          &$(2,1)$   &$(1,2)$   &$-1 $        &$N(2N-1)$ &$1$    \\
$\lb^Y_{\adot}$    &$(1,2)$   &$(1,2)$   &$+1 $        &$N(2N-1)$ &$1$    \\
$q^A$              &$(1,1)$   &$(2,1)$   &$0  $        &$2N$      &$8$    \\
$\eta_\a$          &$(2,1)$   &$(1,1)$   &$-1 $        &$2N$      &$8$    \\
$\etab_{\adot}$    &$(1,2)$   &$(1,1)$   &$+1 $        &$2N$      &$8$    

\end{tabular}
 
It is instructive to
compare the above fields with those of $2N$ 3-branes in type IIB. There
the theory is ${\CN}=4$ $U(2N)$ gauge theory. Type I' $Sp(N)$ theory is
obtained from the ${\CN}=4$ $U(2N)$ theory by $\Omega\cdot Z_2$ projection
(apart from the $O(8)$ fundamental fields which appear from
3-brane-7-brane states and we shall discuss them separately in the
following). Recall that the ${\CN}=4$ theory has an $SU(4)$ R-symmetry
group which can be decomposed in terms of $SO(4)_E \times SO(2)
\equiv SU(2)_A\times SU(2)_Y \times SO(2)$ that appears in the ${\CN}=2$
theory. Writing also $SO(4)_I$ as $SU(2)_{\alpha}\times
SU(2)_{\dot{\alpha}}$ we can write the ${\CN}=4$ fields (all in the
adjoint of $U(2N)$) in terms of their representations under
$SU(2)_{\alpha}\times SU(2)_{\dot{\alpha}} \times SU(2)_A\times
SU(2)_Y \times SO(2)$ as

\begin{eqnarray}
{\rm Vectors} &:& (2,2,1,1,0)\nonumber\\
{\rm Fermions} &:& (2,1,2,1,+1), (1,2,1,2,+1), (2,1,1,2,-1), (1,2,2,1,-1) 
\nonumber\\
{\rm Scalars} &:& (1,1,2,2,0), (1,1,1,1,\pm 2)
\nonumber
\end{eqnarray}
The last entry here denotes the charge under $SO(2)$.

$\Omega\cdot Z_2$ projection acts as follows: $Z_2$ is the center of
$SU(2)_Y$, while $\Omega$ assigns a plus sign to the $Sp(N)$
subalgebra of $U(2N)$ adjoint and assigns minus to the remaining
generators of $U(2N)$ namely the ones transforming as the second rank
anti-symmetric representation of $Sp(N)$.  From this it follows that
vectors are $Sp(N)$ adjoint and transform as $(2,2,1,1,0)$, while the
scalars split into two classes: $\varphi^{AY}$ that transform as
antisymmetric representation of $Sp(N)$ and $\varphi^{\pm}$ transforming
as adjoint. Similarly fermions transforming as $(2,1,2,1,+1)$ and
their complex conjugate $(1,2,2,1,-1)$ are in the adjoint of $Sp(N)$
while the ones transforming as $(1,2,1,2,+1)$ and their complex
conjugate $(2,1,1,2,-1)$ are in the antisymmetric representation.
This is exactly the field content (apart from the fundamentals of
$O(8)$) described above and in fact even their action is
just obtained from the ${\CN}=4$ action by this projection.  This fact
will be important for us in the following because our strategy will
be to use the results of \cite{Dorey} for the case of
$U(2N)$ and project by $\Omega\cdot Z_2$. Not only the zero mode analysis
of \cite{Dorey} can be adapted for our case but even the action
integrals can be carried over from the ${\CN}=4$ analysis. 
   
\subsection{Vector zero modes}

In the $U(2N)$ theory the ADHM data for $k$ instanton is given in 
terms of $(2N+2k)\times 2k$ matrix $\Delta$ of the form \footnote{We use the 
same 
notation as ref. \cite{Dorey}}
\begin{equation}
\Delta_{\l i\adot} = a_{\l i\adot} +
 b_{\l i}^{\a} x_{\a \adot} 
\label{Delta}
\end{equation} 
where $\lambda$ goes over $2N+2k$ indices and the $2k$
indices are split into $i=1,\dots,k$ and $\alpha, \dot{\alpha}=1,2$
and $x_{\alpha \dot{\alpha}}= x_{\mu} \sigma^{\mu}_{\alpha
\dot{\alpha}}$. The matrix $\Delta$ satisfies the constraint
\begin{equation} 
(\bar{\Delta} \Delta)^{\dot{\alpha}}_{ij\dot{\beta}}
= \delta^ {\dot{\alpha}}_{\dot{\beta}}f^{-1}_{ij}
\label{} 
\end{equation} 
where $f_{ij}$ is a symmetric matrix. 
Another ingredient in the construction is $U$ which is a
$(2N+2k)\times 2N$ matrix and satisfies the orthogonality relation
$\bar{\Delta} U = \bar{U}\Delta =0$ and $\bar{U} U =1$.  The
projection operator is $U\bar{U} = 1-\Delta f \bar{\Delta}$.  The
instanton gauge potential $v_{\mu} = \bar{U} \partial_{\mu} U$ gives
a self dual field strength of instanton number $k$. 

We can now go to the $Sp(N)$ case by projecting by $\Omega$ which
implies that $\Delta$ and $U$ satisfy:
\begin{equation}
\Omega_{N+k} \Delta = \Delta^{*} \Omega_k, ~~~~~ \Omega_{N+k} U
=U^{*} \Omega_N
\label{}
\end{equation}
where * denotes complex conjugation and $\Omega_r$ is a $2r\times
2r$ matrix with $r$ diagonal blocks of $\sigma_2$ each.

It is clear that $\Delta$ is defined up to the action of constant $Sp(N+k)$ 
from the left and $GL(k,R)$ from the right. $Sp(N+k)$ acts
simultaneously on $U$ on the left. $U$ on the other hand admits local
$Sp(N)$ transformations on the right. Using the freedom on $\Delta$
we can bring the matrix $b$ in \ref{Delta} to the form
\begin{equation}
b =\left(\begin{array}{c} 0_{2N\times 2k}\\  1_{2k\times 2k}\end{array}\right)
\label{canonb}
\end{equation}
The residual symmetry of $\Delta$ is $Sp(N)$ subgroup of $Sp(N+k)$
left action and $O(k)$ action defined by 
\begin{equation}
\Delta \rightarrow 
\left(\begin{array}{cc} 
1_{2N\times 2N} & 0_{2N\times 2k}\\
0_{2k\times 2N} & g\times 1_{2\times 2} 
\end{array}\right) 
\Delta~ g^t\times 1_{2\times 2}
\label{}
\end{equation}
with $g$ being an $O(k)$ element. This residual $Sp(N)\times O(k)$
symmetry will play an important role in the next section when we
describe the ADHM data using $(-1)$-, 3-, 7-brane system. 

Writing $a= \left(\begin{array}{c} w_{2N\times 2k}\\ a'_{2k\times 2k}\end{array}
\right)$,
the
constraint on $\bar{\Delta}\Delta$ then implies that $a'$ is a
symmetric matrix (i.e. $a'_{(\alpha i)(\dot{\alpha} j)} = a'_{(\alpha
j) (\dot{\alpha} i)}$)  and therefore transforms as the second rank
symmetric tensor under $O(k)$. The constraint also implies that

\be 
D^c_{ij} \equiv tr_2\tau^c(\bar{w} w + \bar{a'} a')_{ ij} =0
~~~c=1,2,3.  
\label{BDterm}
\ee
where $\tau^c$ are Pauli matrices. Note that $D_c$ is a $k\times k$
antisymmetric matrix and transforms in the adjoint representation of
$O(k)$. $D_c$ will play the role of the D-terms in the $O(k)$ gauge
theory of $(-1)$-brane instantons. It is also clear that
$w^{\dot{\alpha}}$ transforms as a bi-fundamental of $Sp(N)\times
O(k)$ and the superscript on $w$ indicates that it transforms as a
doublet of $SU(2)_{\dot{\alpha}}$.  These fields will play the role
of $(-1)$- brane - 3-brane states. 

\subsection{Adjoint and antisymmetric fermion zero modes}

Now let us turn to the fermions. Their zero modes can also be obtained by 
projecting the zero modes for ${\CN}=4$ $U(2N)$ theory. The result for the 
$Sp(N)$ adjoint fermions is
\begin{equation}
(\lambda_{\alpha}^{A})_{uv} = \bar{U}_u^{\lambda}
{\CM}^{A}_{\lambda i} f_{ij} \bar{b}^{\rho}_{\alpha j} 
U_{\rho v} -\bar{U}_u^{\lambda} b_{\lambda  i\alpha} f_{ij} 
({\CM}^T)^{\rho A}_{j} U_{\rho v}
\label{lambdamode}
\end{equation}
where $\CM^{A}$ is a constant $(2N+2k)\times k$
matrix of Grassmann variables. Writing 
\begin{equation}
{\CM}^{A}_{~i} = \left(\begin{array}{c}\mu^{A}_{ui} \\
{\CM'}^{A}_{\beta \ell i}\end{array}\right)
\label{}
\end{equation}
the constraints
on $\CM$ are:
\begin{eqnarray}
(F^{A}_{\adot})_{ij}&\equiv& ({\CM}^T)^{\l A}_i a_{\l \adot j}
+\bar{a}^{\l}_{i \adot}{\CM} ^{A}_{\l j}=0 \nonumber \\ 
({\CM}^{'T})^{A}_{\a} &=& {\CM}^{'A}_{\a}
\label{FDterm}
\end{eqnarray}
In particular this implies that ${\CM}'$ transforms in the symmetric
representation of $O(k)$.

Similarly the zero mode expressions for the fermions $\lambda^Y$ in the  
anti-symmetric representation of $Sp(N)$ 
are given by the above expressions with 
the collective coordinates ${\CM}^A$ 
replaced by ${\CM}^Y$ and $({\CM}^T)^A$
replaced by $-({\CM}^T)^Y$. This in particular implies that ${\CM}^{'Y}$
transforms in the anti-symmetric representation of $O(k)$. 

The fermionic constraints
$F^{A}_{\adot}$ and $F^Y_{\adot}$ transform in the adjoint and the
symmetric representations of $O(k)$respectively. Note that ${\CM '}^A_{\a}$
and $F^A_{\adot}$ carry $(+1)$ charge under $SO(2)$  while  ${\CM '}^Y_{\a}$
and  $F^Y_{\adot}$ carry $(-1)$ charge.
As we shall see in the next section, this is exactly the structure
one finds in the $(-1)$-, 3-, 7- brane
system.

Finally we will need the explicit expressions of the four supersymmetric
and four superconformal exact zero modes. They are given respectively 
by choosing
\be
{\CM}^A_{~ i}=\left(\begin{array}{c} 0 \\ \delta_{\ell i}\psi^A_\b 
                   \end{array}\right) 
\label{ex1}
\end{equation}
and 
\be
{\CM}^A_{~ i}=\left(\begin{array}{c} w_{ui}^{~~\adot}{\xi}^
A_{\adot} \\ 
                    0 \end{array}\right) 
\label{ex2}
\ee

\subsection{Adjoint and anti-symmetric scalars}

In the $U(2N)$ theory the adjoint scalar $\varphi^{\CA \CB}$ transforms
as a vector under the $SO(6)_R$ symmetry. It satisfies the equation of
motion 
\be
D^2 \varphi^{\CA \CB} = \sqrt{2} i [\l^{\CA}, \l^{\CB}].
\label{phieq}
\ee
The solution is given by:
\bea
i\varphi^{{\CA}{\CB}}=-\frac{1}{2\sqrt{2}}{\bar{U}}({\CM}^{\CA}f{\bar{\CM}}^{\CB
}-
{\CM}^{\CB}f{\bar{\CM}}^{\CA})U
+{\bar{U}}\left(\begin{array}{cc} 0_{2N\times 2N} & 0_{2N\times 2k}\\
0_{2k\times 2N} & \Phi^{{\CA}{\CB}}_{k\times k}
\times 1_{2\times 2}\end{array}\right)U
\label{PhiAB}
\eea
where $\Phi^{{\CA}{\CB}}$ is the collective coordinate that satisfies the
equation
\begin{equation}
L\cdot\Phi^{{\CA}{\CB}}=\L^{{\CA}{\CB}}
\label{Lterm}
\ee
with $\L^{{\CA}{\CB}}$ and the operator $L$ defined as:
\bea
L\cdot\Phi^{{\CA}{\CB}}&=&\frac{1}{2}\{ \Phi^{{\CA}{\CB}},W^0\} +
                        [a'_\mu ,[a'_\mu ,\Phi^{{\CA}{\CB}}]] \nonumber \\
\L^{{\CA}{\CB}}&=&\frac{1}{2\sqrt{2}} 
({\bar{\CM}}^{\CA}{\CM}^{\CB}-{\bar{\CM}}^{\CB}{\CM}^{\CA})
\eea
where $W^0 =\tr_2 \bar{w}w$.
Splitting again $SU(4)_R$ symmetry into $SU(2)_A\times SU(2)_Y
\times SO(2)$ and doing the $\Omega\cdot Z_2$ projection, which projects
$\varphi^{AY}$ and $\varphi^{\pm}$ into respectively anti-symmetric and 
adjoint representation of $Sp(N)$ we find that the collective
coordinates $\Phi^{AY}$ and $\Phi^{\pm}$ transform as symmetric
and adjoint representations of $O(k)$ respectively. 
Also as a result of the projection, $\bar{\CM}$ in the above expressions
are replaced by $({\CM}^T)^A$ and $-({\CM}^T)^Y$. 

\subsection{Fundamental fields}

As we mentioned, the theory we are considering also includes
hypermultiplets that transform as bi-fundamental of $Sp(N)\times
O(8)$. In fact we shall be precisely interested in computing the
correlation function of the $O(8)$ flavour currents. The fundamental
fermion zero mode is
\begin{equation}
\eta_u^{\a r}= \bar{U}^\l_u b_{\l i}^{\a} f_{ij} {\CK}_j^r
\label{etamode}
\end{equation}
where $r$ is the $O(8)$ flavour index and ${\CK}$'s are Grassmann numbers
that transform as bi-fundamental of $O(k)\times O(8)$. Note that
$\CK$ does not transform under $SO(4)_I\times SO(4)_E$.

Since the spinor $\eta^{\a r}_u$ couples to the scalar $\varphi^+$ through
the term $\tr(\varphi^+\eta^{\a r}\eta^r_\a)$, then the equation of motion
for $\varphi^+$ has an additional term 
$\eta^{\a r}\eta^r_\a$ on the right hand side
of(\ref{phieq}). The effect of this term is the addition of a term 
\be
(\L_{\rm f})_{ij}={\CK}^r_i {\CK}^r_j
\ee
to the right hand side of equation (\ref{Lterm}) 
but only for the component $\Phi^+$.
In summary, after the projection of the ${\CN}=4~ U(2N)$ result, which contains
those fields of ${\CN}=2~ Sp(N)$ which 
are in the adjoint and antisymmetric reps, the
inclusion of the hypers in the fundamental 
representation is just through the above
modification of the constraint for $\Phi^+$.

\subsection{Multi-instanton measure}

Now we can write down the measure for  $k$-instantons. It is just integration
over the collective coordinates discussed above together with the constraints
(eqns. (\ref{BDterm}), (\ref{FDterm}) and (\ref{Lterm})):
\begin{eqnarray}
&\int & da'~dw~d\mu^A~d\mu^Y~d{\CM}^{'A}~d{\CM}^{'Y}~
d\Phi^{AY}~d\Phi^+~d\Phi^- ~d{\CK}
\nonumber \\ 
& &\delta(F^A_{\adot} ) ~ \delta(F^Y_{\adot} ) 
~\delta (D^c) ~ \delta (L\cdot \Phi^{\CA \CB}-\L^{\CA \CB}_{\rm tot})
~\exp({-\frac{1}{g^2} {\tr}_k\Phi \cdot \L_{\rm tot}})
\label{Int}
\end{eqnarray}
where $\L_{\rm tot}=\L +\L_{\rm f}$. For later use, 
it is convenient to replace the integration over $\Phi^{AY}$
and $\Phi^a$ by
\bea
&\int& d\Phi^{AY}~d\Phi^+~d\Phi^- ~\delta 
(L\cdot \Phi^{\CA \CB}-\L^{\CA \CB}_{\rm tot})
~\exp (-\frac{1}{g^2}{\tr}_k\Phi \cdot \L_{\rm tot})~= \nonumber \\
&\int& db^{AY}~db^+~db^- ~\exp (-{\tr}_k B^{\CA \CB}\cdot L 
\cdot B_{\CA \CB} + \frac{i}{g} {\tr}_k B
\cdot \L_{\rm tot})
\label{bAY}
\eea
Where $B^{\CA \CB}$ is a $4k\times 4k$ 
matrix whose elements  $b^{AY}_{ij}$ and $ b^a_{ij}$ 
 are respectively symmetric and adjoint reps of $O(k)$.

Before analyzing the large $N$ limit of this  measure, it will be 
instructive to make a comparison of the above measure with the one for 
$D(-1)$-branes in the presence of 3-branes and 7-branes. This we will
do in the next section.

\section{ADHM measure and D-instantons}

In the last section, we discussed the ADHM measure for the collective
coordinates of the $k$ instantons in the $Sp(N)$ gauge theory with the specific 
matter content appearing on the type I' 3-branes. 
In this section we will consider the Higgs branch of a system of
$k$ $(-1)$-branes, $N$ 3-branes and 8 of the  7-branes on an orientifold
plane and will see that, as it is expected \cite{W, Doug},
it gives the same measure as the ADHM analysis of the last section. 
The field  content appearing on the world volume of the $(-1)$-branes is

\begin{center}
D$(-1)$-brane  World-volume content
\end{center}
\begin{tabular}{llcccccc}
Bosons &Fermions   &$SO(4)_E$ &$SO(4)_I$ &$SO(2)$ &$SO(k)$             &$Sp(N)$ 
&$O(8)$\\
$b^{AY}$ &         &$(2,2)$   &$(1,1)$   &$0$     &$\frac{1}{2}k(k+1)$ & $1$    
&$1$\\
$b^a$    &         &$(1,1)$   &$(1,1)$   &$\pm 2$ &$\frac{1}{2}k(k-1)$ & $1$    
&$1$\\
$a'_{\a \adot}$&   &$(1,1)$   &$(2,2)$   &$0$     &$\frac{1}{2}k(k+1)$ & $1$    
&$1$\\
  &$\Psi^{\adot Y}$&$(1,2)$   &$(1,2)$   &$+1$    &$\frac{1}{2}k(k+1)$ & $1$    
&$1$\\
  &$\L^{\adot A}$  &$(2,1)$   &$(1,2)$   &$-1$    &$\frac{1}{2}k(k-1)$ & $1$    
&$1$\\
  &${\CM}^{' A}_\a$
                   &$(2,1)$   &$(2,1)$   &$+1$    &$\frac{1}{2}k(k+1)$ & $1$    
&$1$\\
  &${\CM}^{' Y}_\a$
                   &$(1,2)$   &$(2,1)$   &$-1$    &$\frac{1}{2}k(k-1)$ & $1$    
&$1$\\
$w^{\adot}$&        &$(1,1)$   &$(1,2)$   &$0$     &$k$                 & $2N$   
&$1$\\
  &$\mu ^Y$        &$(1,2)$   &$(1,1)$   &$-1$    &$k$                 & $2N$   
&$1$\\
  &$\mu ^A$        &$(2,1)$   &$(1,1)$   &$+1$    &$k$                 & $2N$   
&$1$\\
  &$\CK$           &$(1,1)$   &$(1,1)$   &$-1$    &$k$                 & $1$    
&$8$\\
 
\end{tabular}

Here the fields are classified according to their representations under 
$SO(k)$ gauge theory of the world volume action of the $(-1)$-branes, in which
the $Sp(N)$ and $O(8)$ gauge theories of the 3-branes and 7-branes respectively
are the global symmetries. In addition they carry specific representations of 
the $SO(4)_E \times SO(4)_I \times SO(2)$ which are the rotation groups with E
denoting the 4 directions transversal to 3-branes and longitudinal to 7-branes,
I denoting the 4 directions inside 3-branes world volume and $SO(2)$
denotes the two
transversal direction to the 7-branes. 
As before, the indices $A$ and $Y$ refer to the two
$SU(2)$'s in $SO(4)_E$ and $\a$ and $\adot$ refer to the two $SU(2)$'s
of $SO(4)_I$. It will be useful to make comparison with the field content for 
the $(-1)$-, 3- brane system in type IIB that is relevant for the ${\CN}=4$
theory.
First of
all, we have additional fermionic fields $\CK$ transforming as
bi-fundamentals of
$SO(k)\times O(8)$ arising from open strings between $(-1)$ and 7-branes.
These describe the collective coordinates of the hyper multiplets in the
$(2N,8)$ of $Sp(N)\times O(8)$. Secondly,
the adjoint fields of the $U(k)$ gauge theory in IIB splits into adjoint and
symmetric representations of $SO(k)$ under the $\Omega \cdot Z_2$-projection as
indicated
above. Finally the $(-1)$-3-brane states satisfy a reality condition;
\be
\bar{w}^{\adot}_u = \epsilon^{\adot \bdot} \epsilon_{u v} w^{\bdot}_v
\ee

The $(-1)$-brane action for these fields is 
\be
S=(\frac{1}{g_{-1}^2} S_G+S_K+S_D)
\ee
where
\bea
S_G &=& {\tr}_k \{[b^{AY},b^{BX}]^2 + [b^a,b^{a'}]^2 + [b^{AY},b^a]^2
    +\L^{\adot A}[b_{AY},\Psi_{\adot}^ Y] + 2|D|^2 \}  \nonumber \\
S_K &=& {\tr}_k \{[b^{AY},a'_{\a \adot}]^2 + [b^a,a'_{\a \adot}]^2
               + b^{AY} w^{\adot} _u w^u _{\adot} b_{AY}  
               + b^a w^{\adot}_u w^u _{\adot} b^a
               + {\CM}^{' Y}_\a [b_{AY},{\CM}^{' \a A}]
               \nonumber\\
    & &        + {\CM}^{' Y}_\a [b^-,{\CM}^{' \a}_Y]
               + {\CM}^{' A}_\a [b^{+},{\CM}^{' \a}_A]\}
               + \mu^A b_{AY} \mu^Y + \mu^A b^{-} \mu_A
               + \mu^Y b^{+} \mu_Y + {\CK} b^+ {\CK} \nonumber \\
S_D &=& i {\tr}_k \{[a'_{\a \adot},{\CM}^{'\a A}]\L^{\adot}_A
               +[a'_{\a \adot},{\CM}^{'\a Y}]{\Psi}^{\adot}_Y
               +\mu^{Yu}w^{\adot}_u\Psi_{\adot Y}
               +\mu^{Au} w^{\adot}_u\L_{\adot A} \nonumber \\
& & +D^{\adot}_{\bdot}(w^u _{\adot} w ^{\bdot}_u +
\bar{a}^{'\bdot \a}a'_{\a \adot})\}
\eea

Here $D_{\adot\bdot}$ is an $SU(2)$ triplet of auxiliary fields in
the adjoint of $SO(k)$.
In the infrared limit $\frac{1}{g_{-1}^2}
(=\a'^2) \rightarrow 0$, $S_G$ becomes
irrelevant and therefore we shall drop this term. Note that the remaining 
action $S_D+ S_K$ has the following scaling invariance:
\bea
(w,a')  & \rightarrow & \kappa (w,a') \nonumber\\
(b^{AY},b^a)  & \rightarrow & \kappa^{-1} (b^{AY},b^a) \nonumber\\
(\mu^A,\mu^Y)    & \rightarrow & \kappa^{1/2} (\mu^A,\mu^Y) \nonumber\\
(\L ,\Psi) & \rightarrow & \kappa^{-3/2}(\L ,\Psi) \nonumber\\
({\CM}^{'A} ,{\CM}^{'Y}) & \rightarrow & 
\kappa^{1/2} ({\CM}^{'A} ,{\CM}^{'Y})\nonumber \\
\CK   & \rightarrow & \kappa^{1/2} \CK
\eea
Moreover one can easily see that the integration measure is also invariant 
under the above scaling.
From $S_D$ we obtain the $D$-term constraints. Explicitly, integrating
$D$, $\Psi$ and $\L$ we find the bosonic constraint:
\be
(W^c+(\sigma^c)^{\adot}_{\bdot}~(\bar{a}'a')^{\bdot}_{\adot} )_{[ij]}~ =~ 0 
\label{B-constraint}
\ee
and the fermionic constraints:
\bea
(\mu^Y_u w_{\adot} ^u + [{\CM}^{'\a Y},a'_{\a \adot}])_{(ij)} &=&
0 \nonumber \\
(\mu^A_u w_{\adot} ^u + [{\CM}^{'\a A},a'_{\a \adot}] )_{[ij]}
&=& 0
\label{F-constraints}
\eea
where $W^c$ is defined through the equation
\be
W^0_{ij}\delta^{\adot}_{\bdot} +(\sigma^c)^{\adot}_{\bdot} 
W^c_{ij}=\epsilon^{uv} w^{\adot}_{ui} w _{\bdot vj}
\label{W}
\ee
Note that $W^0$ and $W^c$ are respectively symmetric and antisymmetric
in $ij$.

It is now clear how the correspondence 
between the fields on the $k~(-1)$-branes and
the collective coordinates of the $k$ instantons of previous section goes.
The bosonic fields $a'_{\a \adot}$ and $w^{\adot}$ correspond 
to the two components of the bosonic collective coordinate $a_{\l i \adot}$. 
The other bosonic fields 
$b^{AY}$ and $b^a$ correspond to the coordinates  introduced with the same
notation in (\ref{bAY}). The fermions  
${\CM}^{' A},~ {\CM}^{' Y}, ~\mu^A, ~\mu^Y$ and $\CK$
correspond to the Grassmann coordinates introduced with the same notation in 
section 2  
when we discussed the zero modes of adjoint, antisymmetric and fundamental 
fermions. 
Moreover we see 
that the constraints in (\ref{B-constraint}) and (\ref{F-constraints})
are exactly the same as the ones given in last section and the action $S_K$
for the fields just mentioned is the the same as the action in
(\ref{bAY}). 
   
In order to integrate the delta function constraints (\ref{B-constraint}) and 
(\ref{F-constraints}), we decompose the fermions
$\mu$'s in terms of components $\xi$'s and $\nu$'s parallel and
orthogonal to $w$ respectively:
\bea
\mu^Y_{ui}&=(\xi^Y_{\adot})_{ij} w^{\adot}_{uj} + \nu^Y_{ui} \nonumber \\
\mu^A_{ui}&=(\xi^A_{\adot})_{ij} w^{\adot}_{uj} + \nu^A_{ui}
\label{change}
\eea
The integration over the $2\times 4k(N-k)$ fermionic variables
$(\nu^Y,\nu^A)$ can be easily done since they appear in the action
as $\frac{1}{g}(\nu^A b^{AY} \nu^Y+\nu^A b^- \nu^A+\nu^Y b^+ \nu^Y)$. This then 
gives 
\bea
g^{-4k(N-k)}(det_{4k\times 4k}(B))^{N-k}
\eea

The integration over $4k^2$ out of $2\times (4k^2)$ fermionic variables
$(\xi^Y,\xi^A)$ can be done using the delta-function constraints
(\ref{F-constraints}),
which simply gives $({\det}_{2k\times2k}W)^{2k}$. Note that this factor
cancels exactly  $({\det}_{2k\times2k}W)^{-2k}$ coming from the Jacobian of
the transformations (\ref{change}).
The remaining $\xi$ integrations are over $2k(k-1)$ variables, $\xi^Y$'s, and
$2k(k+1)$ variables, $\xi^A$'s. 

For the integration over $4kN$ bosonic variables $w^{\adot}_{ui}$, as in 
ref.\cite{Dorey}, since the integrand is only dependent on the gauge
invariant matrix $W$, it is convenient to perform the angular 
integrations. This can be done following the method 
explained in \cite{Dorey}. The
steps involve firstly to bring $w$ into a standard upper triangular form
by using the gauge invariance. In our case, using $Sp(N)$ gauge transformation:
\be
w =\left( \begin{array}{ccccc} 
v_{1}^{1} & v_{1}^{2}& .& .& v_{1}^{k}\\
0 & v_{2}^{2} & . & . & v_{2}^{k} \\   
. &.&.&.&. \\ 0&0&.&.& v_{k}^{k}\\ 0&0&.&.&0 \\ ~&~&~&~&. \\ 0&0& .&.&0 
\end{array}\right)
\ee
where the right hand side is an $N\times k$ matrix 
with quaternionic entries and the 
diagonal entries $v_i^i$ are the identity 
elements of the quaternion. This change of
variables and the subsequent angular integration gives the following result:
\be
d^{4kN} w \rightarrow {\rm Vol}(\frac{Sp(N)}{Sp(N-k)}) \prod_{j=1}^k (v_{j}^{j})^{
4N-4j+3} d^{2k(k-1)} v
\ee
where the volume factor appearing above is the product $\prod_{j=1}^k
Vol(S^{4N-4j+3})$. 
Changing further the variables $v$ to gauge invariant variables $W$ defined in 
(\ref{W}) we find that  $\int d^{4kN} w$ is replaced by 
\bea
 {\rm Vol}(\frac{Sp(N)}{Sp(N-k)}) \int d^{k(k+1)/2}W^0 \prod_{c=1,2,3} 
d^{k(k-1)/2}W^c~
                  (\det_{2k\times 2k} W)^{N-k+1/2}
\nonumber
\eea

Moreover the integration over $W^c$ can be done simply be using the bosonic
D-term Delta function which then amounts to replacing 
$W^c$ by $(\sigma^c)^{\bdot}_{\adot}({\bar{a}}'a')^{\adot}_{\bdot}$.

Finally taking into account the factor 
$g^{4Nk+4k}$ coming from the normalization of zero
mode wave functions \cite{Bern}, we are 
left with the following integration measure

\bea
c(g,k,N)\int & d^{k(k+1)/2}W^0~ d^{2k(k+1)}b^{AY}~
d^{k(k-1)}b^a ~
     d^{2k(k+1)}a'~\nonumber\\ 
     & d^{2k(k+1)}{\CM}^{'A} ~
     d^{2k(k-1)}{\CM}^{'Y}~ d^{8k}{\CK}~ 
     d^{2k(k+1)}\xi ^A~ d^{2k(k-1)}\xi ^Y \nonumber\\
     & (\det_{2k\times 2k}W^0)^{N-k+1/2}~
     (\det_{4k\times 4k} B)^{N-k} \exp(-S')
\label{final}
\eea
where
\bea
S'&=&{\tr}_k\{[\BAB,a'_{\a\adot}]^2 + 
\BAB W^0\BAB +\frac{i}{g}\BAB {\Upsilon}^{\CA
\CB}\}
+\frac{i}{g}{\CK} b^+ {\CK} \nonumber \\
c(g,k,N) &=& g^{4k(k+1)} {\rm Vol}(\frac{Sp(N)}{Sp(N-k)})
\label{action1}
\eea
Here 
\be
{\Upsilon}^{\CA \CB}=\zeta^{\CA}W \zeta^{\CB}+
{\mathcal M'}^{\CA}{\mathcal M'}^{\CB} 
\ee
where 

\[\zeta^{\CA}=\left(
\begin{array}{r}
 \xi^Y \\ \xi^A
\end{array}\right)\]

and

\[{\mathcal M'}^{\CA}=\left(
\begin{array}{r} 
 {\CM}^{'Y} \\ {\CM}^{'A}              
\end{array}\right)\]

The above action $S'$, apart from the term
${\CK} b^+{\CK}$, has exactly the same form as the one 
appearing in equation (4.55) of ref. \cite{Dorey}.
The difference is that in the present case,
due to the $\Omega$-projection, the components
of the variables $\BAB$, $\zeta^{\CA}$ and 
${\mathcal M'}^{\CA}$ belong to different representations
of $SO(k)$.
Thus the large $N$, saddle point analysis of \cite{Dorey}
can be adapted to our case, after keeping track of the
above projection. Therefore it will suffice to state the
steps involved and then give the final solution. One first makes the rescaling
$B \rightarrow \sqrt{N} B$ and 
raises
the two determinants in (\ref{final}) to an effective action keeping
only leading terms of order $N$. A class of solution to the saddle point
equations, modulo the $SO(k)$ gauge transformation, is obtained
by setting to zero all the $SO(k)$ adjoint variables ,$b^{\pm}$, with the 
symmetric ones, $W^0, b^{AY}$ and $a'_{\a\adot}$ being diagonal. This
corresponds to the $k$ instantons having independent positions in
$X^i_\mu$ (in $R^4$) and sizes $\rho^i$
with $i=1,\dots,k$ and $\mu=1,\dots,4$,  given
by the diagonal elements of $a'_{\a\adot}$ and $W^0$ respectively. Moreover,
the diagonal elements
of $b^{AY}$ give the positions $\Omega^i_a$ ( $a=1,2,3$ ) of the instantons on
$S^3$.  In other words, each  instanton is parametrized by a point
$(X^i_n, \rho^i, \Omega^i_a)$ of $AdS_5\times S^3$.

Actually in the present case this is not the whole story. We have
also another class of saddle point solutions which correspond to
pairs of instantons splitting off the 7-brane world volume and
sitting at the mirror points in the transverse directions to the 7-
brane. To illustrate this let us consider $k=2r$ be an even number.
in this case we can take $b^{\pm}$ to be diagonal $(r\times r)$ matrix times
$\sigma_2$ and $b^{AY}$,
$a'_{\a\adot}$ and $W^0$ to be diagonal $(r\times r)$ matrix times
$1_{2\times 2}$ subject to the condition 
\begin{equation}
b^{(+} b^{-)} + b^{AY} b_{AY} = (W^{-1})^0
\label{s5}
\end{equation}
If the $r$ eigenvalues are different then the $O(2r)$ gauge group is
broken to the semidirect product $S_r \ltimes (Z_2)^r$ where $S_r$ is
the permutation of the $r$ eigenvalues and $(Z_2)^r$ acts by
reflecting the eigenvalues of $b^{\pm}$ (i.e. $b^{\pm} \rightarrow
-b^{\pm}$). Equation (\ref{s5}), together with the gauge group $S_r
\ltimes (Z_2)^r$ then shows that this saddle point configuration
corresponds to $r$ instantons on $AdS_5 \times S^5/Z_2$ 
(In the covering space
$AdS_5\times S^5$ there are $2r$ instantons at the image points).

Clearly there are also other saddle point solutions intermediate to
the above two cases where some of the instantons stay in the 7-brane
world volume and others split off the 7-brane in pairs. Moreover
some of the eigenvalues may also coincide. But in all cases the
saddle point solution has a clear interpretation of $k$ instantons
forming subsystems of single or "bound" instantons, some of them living in the
7-brane world volume and others living in the bulk 
$AdS_5 \times S^5/Z_2$\footnote{In the case of an ${\CN}=4$ $Sp(N)$
gauge theory the coordinates $(b^{AY}, b^\pm)$ are replaced by
a six-component vector of variables in the antisymmetric of $O(k)$.
The analog of (\ref{s5}) would then give,
together with the other bosonic coordinates, the moduli
space $(AdS_5\times RP^5)^r$ (for $k=2r$), in agreement
with \cite{witt3}.}

\section{Exact saddle points in the large-N limit}

It has been argued in \cite{Dorey}, in ${\CN}=4$ context, 
that, after taking into account the fluctuations
around the saddle point, the leading contribution 
to the integral, in the large $N$ limit, comes from the most degenerate
configuration, corresponding to
the $k$ instantons being of the same
size and in the same position in $R^4$ and $S^5$, i.e.
from the $k$-instanton bound state in $AdS_5\times S^5$. The argument 
rests on the the analysis of the measure when two of the eigenvalues 
become equal. In this limit some of the previously massive variables
become massless. The quadratic determinant of the bosonic variables
therefore diverges in this limit. However in a supersymmetric theory 
the integration over the would-be massless fermionic variables
should cancel this divergence. To see this let us consider $k=2$
case in the ${\CN}=2$ model under consideration (of course the same
consideration will also apply to the ${\CN}=4$ model considered in
\cite{Dorey}). There are two possible generic saddle point
solutions: 
\begin{eqnarray}
&1)& b^{\pm}=0,~~~~{\rm and} ~~b^{AY}, a'_{\a \adot}, W^0~~~{\rm  
diagonal~~with~~ 
different~~eigenvalues},
\nonumber \\
&2)& b^{\pm}=x\sigma_2,~~{\rm and} ~~ b^{AY}, a'_{\a \adot}, W^0~~~{\rm
proportional~~to}~~ 1_{2\times 2}
\label{}
\end{eqnarray}
In the first case the $O(k)=O(2)$ is broken to $Z_2$ while in the
second case it is unbroken. In the first case, to be explicit let us 
assume that the real part of only $b^{AY}$ 
for $A=Y=1$ has different eigenvalues and
all the other variables have equal eigenvalues. Then the off-diagonal
components of $b^{AY}; (A,Y)\neq (1,1)$, $a'_{\a\adot}$,
$b^{\pm}$, 
${\CM}'$'s and $\xi$'s and the imaginary part of $b^{11}$, acquire masses
that vanish in the limit the two eigenvalues coincide (note that a 
certain combination of the off-diagonal components of $W^0$ and other 
fields have quadratic terms that does not
vanish in the limit of coinciding eigenvalues).  Let $M$ be the
difference of the eigenvalues of the real part of $b^{11}$. We would now try to
extract the dependence of the measure on $M$. 
At first sight it may 
appear that there are 9 off-diagonal bosons and only 16 real off-diagonal 
fermions whose masses vanish in this limit, and therefore the
integral over the massive variables diverges as $M^{-9 +\frac{16}{2}}=
M^{-1}$. However by using $O(2)$ transformation one can always
bring the real part of $b^{11}$ into diagonal form, with the Jacobian of the
transformation being $d^3 b^{11} \rightarrow d^2 b^{11} M$ modulo the volume of
$O(2)$,
where on the left hand side the 3 $b^{11}$'s are in the
symmetric representation of $O(2)$ while on the right hand side we
have the 2 diagonal $b^{11}$'s. In the total measure therefore $M$ cancels
as is expected from supersymmetry. 

In the second case, the fact that the measure is independent of the mass 
is even more straightforward to see. $b^{\pm}$, ${\CM}^{'Y}$ and $\xi^Y$
in this case remain massless and the integration over the 
massive bosonic $b^{AY}$ and $a'_{\a \adot}$ and fermionic
${\CM}^{'A}$, $\xi^A$ and ${\CK}$ cancel exactly as expected 
from supersymmetry.

The argument above can be easily extended to other values of $k$ and for 
all the possible saddle point solutions corresponding to subsystems of 
instantons in the 7-brane world volume and in the full 10-dimensions.
It is also easy to see that the same argument applies to the ${\CN}=4$ case 
considered in \cite{Dorey}; there is no divergence in the measure when the 
two eigenvalues coincide. 

What does then select the saddle point in
\cite{Dorey} corresponding to collapse of all the instantons to a
common position and scale? The answer is that in all the other
saddle points there are more fermion zero modes than the 16
corresponding to the supersymmetric and superconformal zero modes.
Indeed
if there are subsystems of instanton 'bound states', there would be
center of mass of each of these subsystems resulting in more fermion
zero modes. In fact these extra fermion zero modes are the ones contained in
${\CM}'$ and $\zeta$ (part of $\mu$ which is parallel to $w$)
in the decomposition of the fermion collective coordinates 
${\CM} =\left(\begin{array}{c}\mu \\ {\CM}'\end{array}\right)$. 
In other words these
extra zero modes are not contained in $\nu$ and $\bar{\nu}$ that are
parts of $\mu$ and $\bar{\mu}$ that are orthogonal to $\bar{w}$ and
$w$ respectively.
If one considers Yang-Mills correlation functions
involving operators that contain only 16 fermions then clearly these
other saddle points cannot contribute. On the other hand if one 
computes correlators involving more than 16 fermions, then as shown
in \cite{Dorey}, the leading order term in large $N$ limit comes when
all the extra fermions (other that 16) appear as $\nu$ and
$\bar{\nu}$ due to the fact that they transform as fundamental of
$SU(N)$ and the $SU(N)$ trace gives a factor of $N$ for each pair
of $\nu$ and $\bar{\nu}$. Thus to the leading order in $N$ only the
saddle point corresponding to all the instantons at same position
and of same scale contribute.

In the ${\CN}=2$ case under consideration the same argument applies. For
any Yang-Mills correlators, the leading order in $N$ arises when one
takes the maximum possible fermions in the $\nu$ and $\bar{\nu}$
part of the fermion collective coordinates. Of course at any saddle
point there are at least 8 exact fermion zero modes coming from ${\CM}'$
and $\zeta$. Precisely 8 exact zero modes appear when all the
instantons sit in the 7-brane world volume at a common position and
with common size. Thus Yang-Mills correlators that contain only 8
fermion collective coordinates $\CM$ (i.e. other than the $O(8)$
fermion collective coordinates $\cal{K}$) will get contribution only
from the saddle point corresponding to the 'bound state' of all the
instantons on the 7-brane world volume. In the next section we will
consider such a correlation function involving four $O(8)$ currents. 
However one may also consider correlators involving more than 8
fermions. If the extra fermions can appear as $\nu$ and $\bar{\nu}$
then the leading order in $N$ is again given by the above saddle point.
But as we shall see in section 5, for correlators involving operators
that couple to Kaluza-Klein modes of closed
string states on $S^5/Z_2$ this leading order in $N$ vanishes, and one
gets contribution from saddle points (for even $k=2r$) that correspond
to $r$ instantons sitting in $AdS_5 \times S^5/Z_2$.
Anticipating these results, in the remaining part of this
section we analyse these two saddle points in detail, the one where
all the instantons sit at a common point in the 7-brane 
and second (for even $k=2r$) when $r$ instantons sit in the bulk 
$AdS_5 \times S^5/Z_2$.

In the first case $b^{\pm}=0$ and all the $b^{AY}$, $a'_{\a \adot}$ and $W^0$
are proportional to identity matrix with the constraint 
$b^{AY} b_{AY} = (W^{-1})^0$.
One can then perform the analysis of small
fluctuations around this configuration and arrive
at the following measure for the collective coordinates 
of $k$ instantons:
\be
N g^4\int \rho^{-5}d\rho d^4X d^3\Omega \prod_{A=1,2}d^2\xi^A d^2\psi^A
{\mathcal{Z}}_k
\label{measure}
\ee
where ${\mathcal{Z}}_k$ is the partition function of the type-I'
$k$ D-instantons in the presence of 8 7-branes at an orientifold
7-plane:
\bea
{\mathcal Z}_k= &\int dX db^a d{\tilde \Theta} d{\Theta}
d{\CK}~\exp(-{\tr}_k([X_i,X_j]^2 + [b^a,X_i]^2 +[b^a,b^{a'}]^2\nonumber\\
&+{\tilde\Theta}[b^-,{\tilde\Theta}] +{\Theta}[b^+,{\Theta}]
+{\Theta}\Gamma^i[X_i,{\tilde\Theta}]) +{\CK} b^+{\CK} ),
\label{dinst}
\eea
where we have assembled $b^{AY}$ and $a'_{\a\adot}$ into an
8-vector $X_i$, $i=1,\dots,8$, of variables in the traceless,
symmetric representation of $SO(k)$, after a rescaling by 
$N^{\frac{1}{4}}$ for each bosonic variable. $\Theta$ 
and $\tilde\Theta$ are chiral $SO(8)$ spinors of opposite chirality 
in the traceless, symmetric and adjoint representation of
$SO(k)$ respectively. They are obtained respectively by
combining $\xi^A$ with ${\CM}^{' A}$  and $\xi^Y$ with
${\CM}^{' Y}$ also taking into account a rescaling factor 
$N^{\frac{1}{8}}g^{\frac{1}{2}}$ for each fermionic variable.
In obtaining the factor $N$ in (\ref{measure}) we have taken into
account a factor $N^{-2kN+k(k-1/2)}$ from ${\rm Vol}(\frac{Sp(N)}{Sp(N-k)})$,
and $N^{-k(k+1)/4}$ from $W^0$ integration.
We should stress that in obtaining the action 
of \ref{dinst}, first one needs to choose a specific basis for the
$SO(8)$ gamma-matrices, given by 
\begin{eqnarray}
\Gamma_{\mu} &=& \gamma_{\mu}\times \sigma_1 \times \sigma_1
~~~~\mu=1,2,3,4 \nonumber\\
\Gamma_5 &=& \gamma_5 \times \sigma_1 \times \sigma_1 ~~~~~~~~~
\Gamma_6 = 1 \times \sigma_2 \times \sigma_1 \nonumber\\
\Gamma_7 &=& 1 \times \sigma_3 \times \sigma_1 ~~~~~~~~~
\Gamma_8 = 1\times 1 \times \sigma_2
\label{gamma}
\end{eqnarray}
Here $\gamma_{\mu}= \left(
\begin{array}{cc}0 & \sigma_{\mu} \\ \bar{\sigma}_{\mu} & 0\end{array}
\right)$
and $\gamma_5=\left( \begin{array}{cc}1 & 0 \\ 0 & -1\end{array}\right)$
are the four dimensional $(4\times4)$ matrices corresponding to
$SO(4)_I$. The first set of sigma matrices corresponds to $SU(2)_R$.
In this basis the 8 dimensional
chirality operator
$\Gamma_9=1\times 1\times \sigma_3$ and the charge conjugation operator is  
$C=c\times \sigma_2 \times \sigma_3$ with 
$c =\left( \begin{array}{cc} \sigma_2 & 0 \\0 &
\sigma_2\end{array}\right)$
being the 4-dimensional charge conjugation operator.  
Secondly, in order to bring the interaction term 
involving $X_i$, $\Theta$ and $\tilde\Theta$
to the form appearing in (\ref{dinst}), one has to perform an
$\Omega$-dependent rotation on the above gamma-matrices.

In the second case (for $k=2r$), $b^{\pm} = M 1_{r\times r} \times
\sigma_2$ and $b^{AY}$, $a'_{\a \adot}$ and $W^0$ proportional to
identity with the constraint eq.(\ref{s5}). After gauge fixing and
integrating over the massive fields, the dependence on $M$
disappears as discussed above, and the action for the
massless variables can again be expanded up to the quartic term. The
resulting action has $U(r)$ gauge symmetry and is given by:
\begin{equation}
\sqrt{N} g^8 \int \rho^{-5}d\rho d^4X d^5\Omega \prod_{A=1,2; i=\pm}d^2\xi_i^A 
d^2\psi_i^A {\mathcal{Z}}^{II}_r
\label{IIdinst}
\ee
where
\bea
{\mathcal Z}^{II}_r= &\int dX db^a d{\tilde \Theta} d{\Theta}
\exp(-\tr_k([X_i,X_j]^2 + [b^a,X_i]^2 +[b^a,b^{a'}]^2\nonumber\\
&+{\tilde\Theta}[b^-,{\tilde\Theta}] +{\Theta}[b^+,{\Theta}]
+{\Theta}\Gamma^i[X_i,{\tilde\Theta}]) ,
\label{dinstII}
\eea
where all the variables are in the adjoint representation of
$SU(r)$ and the $\Gamma$ matrices are the same as in eq.(\ref{gamma}).
Note that in this case there are 16 fermion zero modes in
eq(\ref{IIdinst}) and  the bosonic integral in
(\ref{IIdinst}) can be recognized as the integral over $AdS_5 \times
S_5/Z_2$, with $Z_2$ being the Weyl group acting on the saddle point
solution
$b^{\pm}\rightarrow -b^{\pm}$ subject to the condition (\ref{s5})
defining $S^5$. The quantity ${\mathcal Z}^{II}_r$ in eq.(\ref{dinstII})
is exactly the $r$ type IIB D-instanton matrix theory after having
removed the center of mass variables which is given by 
\begin{equation}
{\mathcal Z}^{II}_r = r^{} \sum_{m|r} \frac {1}{m^2}
\label{}
\end{equation}

\section{Yang-Mills correlators in the instanton background}

After having discussed the ADHM measure and the exact saddle points in 
large $N$ limit, we now compute some Yang-Mills  correlators in the instanton 
background. By the AdS/CFT correspondence, these correlators will be given 
by certain bulk terms in the effective action of the string theory living 
in the bulk. Given the fact that the Yang-Mills correlators will appear 
with the instanton action factor $e^{2\pi i k\tau_{YM}}$ and the relation 
between $\tau_{YM}$ to the complexified string coupling constant $\tau$ in 
type I', the effective action terms that will contribute are precisely the 
ones that get contribution from $k$ D-instantons. It is known via 
heterotic-type I-type I' duality in 8-dimensions, that there are terms
like $t_8 F^4$ and $t_8 R^4$ (and their supersymmetric partners) that 
get contribution 
from world sheet instantons, D-string instantons and D-instantons in these 
three dual theories respectively. These terms live in the 7-brane world 
volume in the type I' theory. However we will argue that there are
also some bulk terms in the type I' effective action (i.e. living in
10-dimensions) of the form $t_8 t_8 R^4$ and its supersymmetric partners
that receive corrections from D-instantons. We will first compute some 
correlators in Yang-Mills that in the large $N$ limit receive contribution 
from the saddle 
point corresponding to all instantons sitting on the 7-brane world
volume and show that they can be reproduced as coming from the $t_8
F^4$ and $t_8 R^4$ terms in the 7-brane bulk theory. Next we will
consider some correlator that selects the saddle point which
corresponds to instantons sitting in the bulk $AdS_5\times S^5/Z_2$
and show that they come from the $t_8 t_8 R^4$ term in the string
theory.
 
\subsection{$O(8)$ current correlators and $AdS_5$ propagators}

The $O(8)$ gauge 
potential couples to the flavour current in
the 3-brane world volume theory via the term $A_{\mu}^a J_{\mu}^a$ 
with $J_{\mu}^a =\bar{q} T^a \overrightarrow{D}_{\mu} q
-\bar{q} T^a \overleftarrow{D}_{\mu} q
+ \bar{\eta}^{\a A}T^a \sigma_{\mu \a \adot} \eta_A^{\adot}$ 
and $D_{\mu}= \partial_{\mu}+ A_{\mu}$ is the $Sp(N)$ 
covariant derivative. On the Yang-Mills side we are interested in
computing a 4-point function of the currents $J_{\mu}^a$ in the
presence of $Sp(N)$ instantons.  We will see below that this is sufficient 
to soak the 8 exact gaugino zero modes and moreover for odd instantons will 
soak also the 8 fundamental
zero modes giving rise to odd parity $O(8)$ invariant. 
In the large $N$ saddle point approximation, the
correlation function is just given by plugging in the classical solutions 
of the
fields. First let us consider the bosonic part of the $O(8)$ current.
The scalar field $\phi$ satisfies the equation:
\begin{equation}
D^2 q = \lambda \eta
\label{}
\end{equation}
where $\lambda$ and $\eta$ are the gaugino and fundamental fermion fields. 
Plugging
in the expressions for the  zero
modes of $\lambda$ and $\chi$ as given in equations (\ref{lambdamode}) and 
(\ref{etamode}) in terms of the ADHM data, we 
can solve the above equation with the result:
\begin{equation}
q^{r}_u = \bar{U}_u^{\rho} {\CM}_{\rho i} f_{ij}{\CK}_j^r
\label{q}
\end{equation}
where the index $r$ and $u$ label the fundamentals of $O(8)$ and $Sp(N)$
respectively, $\rho$ runs over $2N+2k$ values, and $i,j$ run from 1 to $k$.
 
Using now the explicit form for ${\CM}$ for the case of 
the 8 exact zero modes for the
gauginos (\ref{ex1}),(\ref{ex2}), and the saddle point solution for which 
\[ \begin{array}{ll}
f_{ij}= \frac{1}{y^2+\rho^2} \delta_{ij}~~~~~~~~~~~~
&(a'_{\mu})_{ij}= -X_{\mu} \delta_{ij}   \\
(W^0)_{ij} = \rho^2 \delta_{ij}          &W^c = 0 \end{array} \]
where $y=x-X$, we can express the contribution of the scalars to the 
current $J_{\mu}$, after some straightforward 
 algebra, as
\begin{eqnarray}
J_{\mu} = \frac{1}{N^{1/4}g} \rho^2 f^4 
{\CK} T^a {\CK} &\big[ {\psi}^{\alpha A} 
{\psi}_{\beta A} 
y_{\nu} (\sigma^{[\nu} \bar{\sigma}^{\mu ]})_{\alpha}^{\beta} +
\xi_{\dot{\alpha}}^A
\xi^{\dot{\beta}}_A \rho^2 (\bar{\sigma}^{[\mu}
\sigma^{\nu]})^{\dot{\alpha}}_{\dot{\beta}} 
y^{\nu}
\nonumber \\ &+ 2{\psi}^{\alpha A}\xi^{\dot{\beta}}_A ((y^2-\rho^2)
\sigma^{\mu} - 2 y^{\mu}y^{\nu} \sigma^{\nu}) _{\alpha 
\dot{\beta}} \big]
\label{jbose}
\end{eqnarray}

The fermion part of the current $\bar{\eta}^{\alpha A}T^a 
{\sigma_{\mu}}_{\alpha \adot} \eta_A^{\adot}$ is a bit more tricky.
At first sight it appears that this term cannot soak the 8 exact 
supersymmetric and superconformal zero 
modes. Indeed the zero mode of $\eta^{\alpha}$
does not contain ${\CM}^A$ collective coordinates and $\bar{\eta}^{\adot}$
has no zero mode of the Dirac operator.
However a closer inspection shows that one needs to solve the
classical equation for $\bar{\eta}^{\adot}$ in the presence of
the 8 exact zero modes,
\be
D_{\alpha \adot} \bar{\eta}^{\adot} = (\lambda_{\alpha}^A q_A +
\varphi^{+} \eta_{\alpha})
\label{etabareq}
\ee
Note that this equation does not have a solution for arbitrary 
collective coordinates ${\CM}^A$. The reason for this is that the right
hand side of the above equation has a non-zero overlap with the
zero modes for $\eta^{\alpha}$ given in (\ref{etamode}) i.e. zero modes of
the conjugate operator $\bar{D}_{\adot \alpha}$). This is evident from
the $N=2$ analogue of the equation (3.7) of \cite{Dorey} which reads
\be
D_{\alpha \adot} \bar{\eta}^{\adot} = (\lambda_{\alpha}^A q_A +
\varphi^{+} \eta_{\alpha}) + \chi_{\alpha}
\label{etabareq1}
\ee
where
\be
\bar{\eta}^{\adot} = \frac{1}{2}\bar{U} {\CM}^A f \bar{\Delta}^{\adot} {\CM}_A f 
{\CK}
                    +\frac{1}{2}\bar{U} \hat{\Phi}^{+} {\Delta}^{\adot} f 
{\CK} 
\label{etabar}
\ee
and
\be
\chi_{\alpha} = \bar{U} b_{\alpha} f \Phi^{+} \CK
\ee
is a zero mode of the operator $\bar{D}^{\adot \alpha}$ and
$\hat{\Phi}^+=\left(\begin{array}{cc} 0 & 0\\0 & \Phi^+\times 1_{2\times2}
\end{array}\right)$ is the collective coordinate of $\varphi^{+}$ appearing
in (\ref{PhiAB}). It is clear therefore that $\chi_{\alpha}$ is not in the
image of the operator $D_{\alpha \adot}$.

A careful analysis of the collective coordinate approach followed
here
(as well as in the previous papers \cite{Dorey}, and others) shows
that
in the correlator, in the leading semiclassical approximation
$\bar{\eta}^{\adot}$ should be replaced by the right hand side of
eq.(\ref{etabar}). However here we will give a simple argument to
prove this which is based on the fact that we need to soak the 8
exact zero modes.

The right hand side of the expression in terms of collective
coordinates is quadratic in ${\CM}^A$'s. Since we need 8 exact zero
modes and we have only 4 currents, it follows that each of these
${\CM}^A$'s must be replaced by the exact zero modes. Therefore it
suffices to solve eq.(\ref{etabareq}) in the presence of
exact zero modes. In this case using eq.(\ref{Lterm})
one finds $\Phi^+=0$. This is due to the fact that
${\CM}^{A} {\CM}_A=0$ for exact zero modes and the operator $L$ 
is a positive definite operator. As a result $\chi_{\alpha}$ is zero
and eq.(\ref{etabar}) solves eq.(\ref{etabareq}).

One can now, using eqs.(\ref{etabar}),(\ref{etamode}), 
compute the fermionic part of
the $O(8)$ current for the 7-brane saddle point under consideration.
The result turns out to be identical to the bosonic part of the
current, namely the right hand side of (\ref{jbose}). Thus the total
$O(8)$ current is proportional to (\ref{jbose}). 
It is also easy to check that the current is conserved 
($\partial_{\mu} J_{\mu} =0$).

The 8 exact gaugino zero modes can be grouped 
into an 8 component chiral spinor of 
8 dimensional Euclidean space as
\begin{equation}
S = \left(\begin{array}{c} {\psi}_{\alpha 1}\\  \rho
\xi^{\dot{\alpha}}_1 \\ {\psi}_{\alpha 2} \\ 
\rho \xi^{\dot{\alpha}}_2 \end{array}\right)
\end{equation}
Furthermore we will use the 8 dimensional $(16 \times 16)$ $\Gamma$ matrices 
introduced in (\ref{gamma}).
Note that $\gamma_\mu$ and $\gamma_5$  act on $SO(4)_I$ 
indices $\alpha$ and $\dot{\alpha}$, 
and that, from the two sets of sigma matrices used,
the first one
acts on the $SU(2)_R$ index $A,B$ while the second one acts on the
two chiralities of 8-dimensional Euclidean space.  Moreover the
charge conjugation matrix $C$, as defined in section 4, raises and lowers
the indices $\alpha$, $\dot{\alpha}$ and $A$. 
The current $J_{\mu}^a$ now becomes:
\begin{equation}
J_{\mu}^a(x) =\frac{1}{N^{1/4}g} \sum_{m,n=1}^5 S^T C 
(1+\Gamma_9) \Gamma^{mn} S~{\CK}T^a {\CK}~
\partial_{[m} G_{n] \mu}(X,\rho;x)
\label{}
\end{equation}
where $G$ is the propagator of the gauge potentials on the $AdS_5$ space with
$z=(X,\rho)$ being identified with a point in $AdS_5$ 
bulk \cite{Witten, Freedman}
\be
G_{n \mu} (X,\rho;x) = \frac{3}{\pi^2}\frac{\rho^2}{\rho^2+(X-x)^2} 
J_{n\mu}(z-x)
\label{}
\ee
where $J_{nm}(z)=\delta_{mn}-\frac{2z_m z_n}{z^2}$.
Inserting the four $O(8)$ currents and integrating over the 
8 exact gaugino zero modes 
produces the $t_8$ tensor and we
get the result:
\bea
< \prod_{i=1}^4 J_{\mu_i}^{a_i}(x_i) >_{YM}= \int d^4 X
\frac{d\rho}{\rho^5} d^3 \Omega & \prod_{i=1}^4 <A_{\mu_i}^{a_i}(x_i) 
F_{m_i,n_i}^{b_i} (X,\rho)>_{AdS_5}\nonumber\\ 
&\times t_8^{m_1 n_1 m_2 n_2 m_3
n_3 m_4 n_4}
Z^{(k)}_{b_1 b_2 b_3 b_4}
\label{}
\eea
where 
\begin{equation}
Z^{(k)}_{b_1 b_2 b_3 b_4} = < \prod_{i=1}^4 {\CK}T^{b_i}{\CK} >_{D-inst}
\label{kk}
\end{equation}
Using either the computation of the D-string instantons in type I via the Matrix
String approach or by using the 
heterotic world-sheet instanton result, which will
be discussed in the next section,  we conclude
that 
\begin{eqnarray}
Z^{(k)}_{b_1 b_2 b_3 b_4} &=& e^{2\pi i k \tau} \epsilon^{i_1 \dots i_8}
J^{b_1}_{i_1
i_2} J^{b_2}_{i_3
i_4} J^{b_3}_{i_5 i_6} J^{b_4}_{i_7 i_8}\sum_{\ell|k} \frac{1}{\ell}  ~~~~
{\rm for~ k~ odd}
\nonumber\\
Z^{(k)}_{b_1 b_2 b_3 b_4} &=& e^{2 \pi i k \tau} 
[{\tr} (J^{(b_1} J^{b_2} J^{b_3}
J^{b_4)})C_k^{F^4}
\nonumber \\
&~& + {\tr} (J^{(b_1} J^{b_2}) tr (J^{b_3} J^{b_4)}) C_k^{F^2F^2}  ~~~~~~~
{\rm for~ k~ even}
\label{zk}
\end{eqnarray}
where

\[ C_k^{F^4}=\left \{ \begin{array}{ll}4\sum_{\ell|k} \frac{1}{\ell}
 & \mbox{for $k=4m-2$}\\
4\sum_{\ell|k} \frac{1}{\ell}-4\sum_{\ell|(k/2)} \frac{1}{\ell} 
& \mbox{for $k=4m$}\end{array}\right. \]
and

\[ C_k^{F^2F^2}=\left\{ \begin{array}{ll}2\sum_{\ell|k} \frac{1}{\ell}
& \mbox{for $k=4m-2$} \\
2\sum_{\ell|k} \frac{1}{\ell}-\sum_{\ell|(k/2)} \frac{1}{\ell}
& \mbox{for $k=4m$}\end{array}\right. \]

In (\ref{zk}) $J$'s are the $O(8)$ generators 
in the vector representation, $\tau=a +i/g_{str}$
with $a$ being the type I' 0-form RR potential and in the second equation on the
right hand side $b_i$'s are
totally symmetrized. Note that the $k$ odd 
term above gets contribution only from
the $O(8)$ odd parity invariant. This could be seen directly at the level of
D-instanton action. Indeed the the $O(8) \times O(k)$ fermions ${\CK}$ couple
only to 
the $O(k)$ gauge fields $b^{+}$ in the D-instanton 
action. Integrating the ${\CK}$
therefore gives $(\det b^{+})^4$ in the 
vector representation of $O(k)$. For odd $k$,
$b^{+}$ has always at least one zero mode for generic values of $b^{+}$. Thus 
to get a non zero result one has to soak the fermion zero modes, 
which is acomplished by the insertions of 8 ${\CK}$
appearing in eq.(\ref{kk}), arising from 
the four $O(8)$ currents. Soaking these
zero modes brings about the 8-rank $\epsilon$ tensor in eq.(\ref{zk}) for odd
$k$.

So far in the above discussion, the $S^3$ part of the measure has played no
role apart from giving an overall volume factor. This is because we were
considering correlators of the $O(8)$ currents which couple on the
$AdS_5$ side to the lowest Kaluza-Klein modes of the $O(8)$ gauge fields
on $S^3$. There are two types of higher Kaluza-Klein modes one can
consider: that of the closed string states such as dilaton, graviton etc.
and that of the 7-brane world volume fields namely $O(8)$ gauge fields.
For the closed string fields Kaluza-Klein modes will be on $S^5/Z_2$
while for the 7-brane fields these modes will be on $S^3$
and therefore one would expect the integral to be on the 7-
brane world volume namely $AdS_5 \times S^3$. In particular 
the correlators involving Yang-Mills operators that couple to the 
Kaluza-Klein modes coming from the 7-brane fields, will get contribution 
only from 7-brane saddle point. These correlators can be calculated 
exactly as in the case of 
$N=4$ theory discussed in \cite{Dorey} and one can show that they probe the
$S^3$ part of the space-time.

\subsection{Correlators related to bulk $R^4$ terms}

We will now consider a correlator which will get contribution from
the saddle point with $b^{\pm}\neq 0$ that probes the full
$AdS_5\times S^5/Z_2$. 
It is clear that such correlators must involve operators
that do not carry $O(8)$ quantum numbers. Such operators can therefore be
obtained by projection from the ${\CN}=4$ operators. In the ${\CN}=4$ context 
the operators $O^{ab} = tr \varphi^a \varphi^b$ in the (20) of $SO(6)$ (i.e.
traceless
symmetric tensor of 2 $SO(6)$ vectors, with $a,b$ here labelling the $SO(6)$
vectors) carry dimension 2, and they couple via AdS/CFT correspondence to 
a certain Kaluza-Klein mode $J$ of the scale factor $h_{\alpha}^{\alpha}$ and
the 4-form potential $C_{\alpha \beta \gamma \delta}$ on $S^5$. The combination
$J$ of these two fields given by
\be
J(x) = \int_{S^5} (h^{\alpha}_{\alpha} + \epsilon^{\alpha \beta \gamma \delta 
\rho} C_{\alpha \beta \gamma \delta} D_{\rho}) Y^{(2)}
\ee
where $Y^{(2)}$ is the second spherical harmonic on $S^5$. In other words the
relevant expansions of $h_{\alpha}^{\alpha}$ and $C_{\alpha \beta \gamma 
\delta}$
are
\be 
h^{\alpha}_{\alpha} = J(x) Y^{(2)}, ~~~~~~~~C_{\alpha \beta \gamma \delta}=
J(x)\epsilon_{\alpha \beta \gamma \delta \rho} D^{\rho} Y^{(2)}
\ee
The projection to ${\CN}=2$ splits (20) of $SO(6)$ into $(3,3)_0$,
$(1,1)_{\pm2}$
and $(1,1)_0$ of $SU(2)_A \times SU(2)_Y \times U(1)$, the representations
$(2,2)_{\pm 1}$ being projected out. On the Yang-Mills side these operators 
are respectively ${\tr} \phi^{AY} \phi^{BX}$ 
(where $A,B$ and $X,Y$ are symmetrized), 
${\tr} (\phi^{\pm})^2$ and ${\tr}\phi^+
\phi^-$. On the AdS side these operators couple to field $J$
mentioned above, where the spherical harmonics 
$Y^{(2)}$ is restricted to even ones
under the $Z_2$ projection. 

The
simplest correlator is the one involving four operators
$O_{(3,3)}={\tr}\phi^{AY}\phi^{BX}$ with $A,B$ and $X,Y$ symmetrized.
In the large-N limit, we must replace the fields by their classical
solution with the result:
\be
O_{(3,3)} = \frac{1}{g^2}{\tr} \bar{U} ({\CM}^A f {\CM}^{TY}+{\CM}^Y f
{\CM}^{TA})  U \bar{U} {\CM}^B f {\CM}^{TX} U
\label{}
\ee
The correlator then would involve 8 ${\CM}^A$'s which will soak the 8
exact zero modes, leaving behind 8 ${\CM}^Y$'s. 
This correlator gets contribution from both the saddle
points described in the previous section:

1) When all the instantons sit at the same point in the 7-brane
world volume (i.e. $b^{\pm}=0$). In this case
using the fact that $\nu^Y$ appear in the action in the combination
$\nu^A \nu^Y b_{AY} + \nu^Y \nu_Y b^+$ and the fact that $b^+=0$
in this saddle point, we conclude that all the 8 ${\CM}^Y$'s give rise
${\CM}^{'Y}$'s and $\zeta^Y$'s.
The scaling of $({\CM}')^Y$ and
$\zeta^Y$ by $\sqrt{g}N^{-1/8}$ and the measure factor $Ng^4$
in eq.(\ref{measure}) gives rise to a term which goes as order one in $g$ and
$N$  times an
$O(2r)$ Matrix Theory correlator of 8 $O(2r)$ adjoint fermions
in the matrix theory integral. In the supergravity theory
this should correspond to 7-brane $t_8 R^4$ and $t_8 (R^2)^2$ terms
together with the terms involving the 4-form potential. Such terms via
T-duality and S-duality should be obtainable 
from the heterotic one loop 
computation involving gravitons and anti-symmetric tensor fields.

2) When $r$ instantons sit at a common point in $AdS_5 \times S^5/Z_2$
(with the other $r$ instantons sitting at the image point) there are 
8 zero modes of ${\CM}^Y$'s as well that are not lifted and the resulting 
matrix theory is that of $SU(r)$ theory of type IIB instantons. Thus
the result should be the same as in the type IIB case where this correlator
is expected to get contribution from $t_8 t_8 R^4$ (and terms involving 4-form
potential). The measure factor $\sqrt{N}$ can be understood as follows.
The 10-dimensional integral of $R^4$ goes as $L^2 \sim \sqrt{g^2 N}$ with $L$ 
being the radius of
$S^5$. The factors of $g$ cancel due to the fact that the leading D-instanton
contribution to $R^4$ goes as $1/g$ \cite{gg, Banks, Bianchi}. Note that in
type I'
theory, the 0-form RR field $\chi$ is not projected out and therefore the
10-dimensional type IIB
D-instanton contributions would survive in type I' theory.

\section{String theory computation of $F^4$ terms}

In this section we discuss the $F^4$ computation for $SO(8)^4$
heterotic theory in 8 dimensions. The world sheet instantons of the
heterotic theory
are then mapped via S-duality to D-string instantons in type I theory,
which in turn are mapped via two T-dualities to $(-1)$-brane D-instantons 
in type I' theory. In the present context of the near horizon limit
of the 3-branes lying at an orientifold plane where 8 of the 7-
branes live, we are interested in all the four $F$'s in the same $SO(8)$
factor. There are three independent quartic invariants of $SO(8)$:
\begin{equation}
\tr F^4,~~~~~~ (\tr F^2)^2, ~~~~~~~ \epsilon^{i_1 \dots i_8}
F_{i_1 i_2} F_{i_3 i_4} F_{i_5 i_6} F_{i_7 i_8}  
\end{equation}
Here tr denotes trace in the vector representation of $SO(8)$ and
$\epsilon$ is the 8-rank anti-symmetric tensor in the vector
representation, $F_{ij} \equiv F_a (J^a)_{ij}$ with $J^a$ being the
$SO(8)$ generators in the vector representation. Note that the first two 
invariants
are $O(8)$ parity even while the third one is $O(8)$ parity odd due to the
$\epsilon$ tensor. The amplitudes involving the first two
invariants, namely $\tr F^4$ and $(\tr F^2)^2$, have already been
computed in \cite{Lerche, gut}, with the result
\begin{eqnarray}
\Delta^{F^2F^2}&=&4[\log|\eta(4T)|^4-2\log|\eta(2T)|^2]-
4\log[T_2U_2|\eta(U)|^4],\nonumber\\
\Delta^{F^4}&=& -\frac{1}{2}[\log|\eta(4T)|^4-\log|\eta(2T)|^2]
\label{thr}
\end{eqnarray}
Here $T$ is the complexified Kahler modulus of the torus $T= B_{12}+
i R_1 R_2$. The coefficient 
of $e^{2\pi i n T}$ in the above denotes the contribution of $n$ world sheet 
instantons to the amplitude. Note that there are only even number of
instantons contributing to these two $SO(8)$ invariants. We shall
now compute the last invariant, namely the $O(8)$ parity odd
invariant. We will see that only odd instantons contribute to this
term.

$SO(8)^4$ theory is best seen as a $Z_2 \times Z_2$ orbifold
of heterotic $SO(32)$ theory compactified on a torus of radii $2R_1$
and $2R_2$. The orbifold group is generated by two elements $g$ and
$h$ where $g$ is half shift along first direction together with 
the Wilson line $(sp,sp,0,0)$ in the decomposition of $SO(32)$ in
terms of $SO(8)^4$ and $h$ is the half shift along the second
direction together with the Wilson line $(sp,0,sp,0)$. Here $sp=(\frac{1}{2})^4$
denotes the highest weight of the spinor representation of $SO(8)$. 
Including all
possible twists along $\sigma$ and $t$ directions of the world-sheet
torus, we find 16 sectors which split into 5 modular orbits: one
containing the completely untwisted sector, 3 orbits containing 3
sectors each and one orbit containing 6 twisted sectors. Denoting by $(a,b)$
the twisted sectors with twist $a$ along $\sigma$ and $b$ along $t$ 
directions, we can represent these orbits as $(1,1)$ for the completely
untwisted sector, $(1,g)$, $(1,h)$ and $(1,gh)$ for the three orbits
containing 3 sectors each and finally $(g,h)$ for the orbit
containing 6 sectors. The first 4 orbits contribute to the $O(8)$ even
parity invariants and they have been already studied in \cite{Lerche, gut}.
The last
orbit represented by $(g,h)$ contributes to the $O(8)$ odd parity
term. 

The instanton numbers of course are governed by the windings along
$\sigma$ and $t$ directions. Let us denote by $(n_1,n_2)$ the windings of 
$(X_1,X_2)$ along $\sigma$ direction and $(m_1,m_2)$ that along
$t$ direction. World sheet instanton action is $(2R_1)(2R_2) 
|n_1m_2-n_2 m_1|$.
Due to shifts associated with $g $ and
$h$ the windings in these representatives of the 5 modular orbits
are:
\begin{eqnarray}
(1,1)&:& (n_1,n_2; m_1,m_2) \nonumber\\
(1,g)&:& (n_1,n_2, m_1+\frac{1}{2}, m_2) \nonumber \\
(1,h)&:& (n_1,n_2; m_1,m_2+\frac{1}{2}) \nonumber\\
(1,gh)&:& (n_1,n_2; m_1+\frac{1}{2},m_2+\frac{1}{2}) \nonumber\\
(g,h)&:& (n_1+\frac{1}{2},n_2; m_1,m_2+\frac{1}{2}) 
\label{}
\end{eqnarray}
Instanton number being invariant under the modular group, it is
clear that in the first 4 orbits the instanton action is even number
times $R_1.R_2$ while in the last orbit it is odd number times
$R_1.R_2$.

Let us consider the $SO(32)$ fermion partition function. 
In the $(g,h)$ sector, we have a twist along $\sigma$ direction by
$(sp,sp,0,0)$ and along $t$ by $(sp,0,sp,0)$. This means that the
characteristics of theta functions shift (in groups of four) by 
$(1/2,1/2)$, $(1/2,0)$, $(0,1/2)$ and $(0,0)$ respectively. Now in
the original $SO(32)$ theory one has the sum over all spin
structures for all the 32 fermions simultaneously. The shifts in the
characteristics 
mentioned above imply that the partition function now becomes:
\begin{eqnarray}
\theta_3^{16} +\theta_4^{16} + \theta_2^{16} \pm \theta_1^{16}
\rightarrow &\theta_1^4 \theta_2^4 \theta_4^4 \theta_3^4
+\theta_2^4 \theta_1^4 \theta_3^4 \theta_4^4
+ \nonumber \\ &\theta_4^4 \theta_3^4 \theta_1^4 \theta_2^4
+\theta_3^4 \theta_4^4 \theta_2^4 \theta_1^4
\end{eqnarray}
This partition function is of course zero because of the appearance
of $\theta_1^4$ in each term on the right hand side. However since
$F^4$ vertex operators contain 4 Kac-Moody currents, they can in 
principle give non-zero answer. Indeed if we take for instance
the four Kac-Moody currents to be in four different Cartan directions of 
a given $SO(8)$, then one of the terms in the right hand side will 
contribute since the effect of introducing these currents is to take
derivatives of the appropriate $\theta(\tau,z)$ with respect to $z$
at $z=0$. The result therefore becomes:
\begin{equation}
{\theta'}_1^4 \theta_2^4 \theta_3^4 \theta_4^4 = 2^8 \pi^4 \eta^{24}
\end{equation}
Remarkably $\eta^{24}$ cancels the $1/\eta^{24}$ coming from the
partition function of the oscillator modes in the bosonic string
sector (the right moving sector) of heterotic string. Left moving 
part of the heterotic string, i.e. the fermionic string sector, gives 
the usual $t_8$ tensor in the presence of 4 gauge field vertices as can be
easily seen in the Green-Schwarz formalism. Indeed each of the $F$ vertex 
contains two Green-Schwarz fermions that soak the 8 fermion zero modes in 
the light cone gauge and all the non-zero mode determinants cancel
due to space-time supersymmetry. Thus the final result for the
coefficient $Z(T,U)$ of the $O(8)$ odd parity term
$\epsilon^{i_1 \dots i_8}
F_{i_1 i_2} F_{i_3 i_4} F_{i_5 i_6} F_{i_7 i_8}$ 
is just given by the 2-dimensional lattice sum of the windings.:
\begin{equation}
Z(T,U) =
\int_{\cal{M}} \frac{d^2\tau}{\tau_2^2}4T_2 \sum_M e^{8\pi i |det M|T}
e^{-\frac{4\pi T_2}{\tau_2 U_2}\big |(1,U)M\big ({\tau \atop -1}\big )\big | ^2}
\label{}
\end{equation}
where 
\be
M= \left(\begin{array}{cc} n_1 + \frac{1}{2} & m_1\\  n_2 & 
m_2 + \frac{1}{2} \end{array}\right),
\ee
with $n_i$ and $m_i$ being integers.
$U$ is the complex structure of the torus and 
$\mathcal{M}$ is the fundamental domain of the $\Gamma_2$ subgroup of
$SL(2,Z)$ defined as the set of matrices $\left(\begin{array}{cc}
a&b\\c&d\end{array}\right)$ with $b$
and $c$ even integers. Note that $\mathcal{M}$ has 6 copies of the
fundamental domain of $SL(2,Z)$ owing to the fact that the orbit of
$(g,h)$ consists of 6 twisted sectors. One can now use suitable
elements of $\Gamma_2$ to set $n_2=0$ and the result is the
unfolding of the domain $\cal{M}$ to the full upper half plane. The
integration
over the upper half plane yields the result:
\begin{equation}
Z(T,U)= 2(\log|\eta(T)|^2)_{odd}
\label{odd}
\end{equation}
where the subscript $odd$ indicates projection onto odd powers of
$e^{2\pi i T}$. Notice that there is no $U$ dependence since in this
sector there is no degenerate orbit.

Equations (\ref{thr}) and (\ref{odd}) give the complete $F^4$ terms
in the $SO(8)^4$ heterotic theory in 8-dimensions. In type I also
one can do
an analogous calculation involving D-string instantons wrapped on
the torus following the methods of \cite{bachas,gmnt}. The basic idea
here is that in the infrared limit, the theory of $N$ D-strings
collapses into a symmetric product of $N$ copies of heterotic string
(in the static gauge). The instanton contribution is obtained by
considering the one loop amplitude involving 4 $F$ vertices with
$\sigma$ and $t$ being identified with $X_1$ and $X_2$ respectively,
as is dictated by the static gauge. We will not give the details
here but a trivial extension of the results of \cite{bachas,gmnt} after
including the Wilson lines in $g$ and $h$ yields the same result as 
the heterotic string one loop result. Finally in the type I' theory
which is obtained by 2 T-dualities on type I theory, the coupling
$F^4$ should be the same as in type I, however a direct computation
of D-instanton contributions in type I' theory is an interesting open
question.

\section{Conclusions}

In this paper we studied the multi-instanton 
effects in ${\CN}=2$ $Sp(N)$ Yang-Mills theory
that appears in the 3-brane world volume in the type I' 
theory (2 T-dual of type I).
We showed that the ADHM construction of the multi-instantons gives rise to 
collective coordinates which can be interpreted as the fields living in the
$(-1)$-brane instantons in the presence of 3-branes in type I'. 
The saddle points in
the large N-limit describes $(-1)$ brane instantons sitting at various
points in the
7-brane world volume ($AdS_5 \times S^3$) or in 
the 10-dim bulk ($AdS_5 \times S^5/Z_2$) that
appears as the near horizon limit of the 
3-brane solutions. We further discussed the
nature of the exact saddle points in the large-$N$ 
limit and argued that
there are 
two distinct classes of exact saddle points, 
one where all the instantons sit at the
same point in the 7-brane world volume and the other when all the instantons
sit at a common point in the 10-dim bulk. 
The latter is of course possible only for
even number of instantons since there has 
to be an equal number of instantons at a
given point and its image under the $Z_2$ action. 
While the 7-brane saddle point
measure is order 1 in $N$, the bulk saddle 
point measure goes as $\sqrt{N}$. 

In the 7-brane saddle
point there are 8 exact fermion zero 
modes and the measure splits into an integral
over $AdS_5 \times S^3$ times a
matrix integral over the $(-1)$ brane world volume theory ($O(k)$ gauge
theory) that appears in type I' where the center 
of mass variables have been factored out. 
This matrix integral via 2 T-dualities 
is related to D-string instanton integrals
in type I (which can be evaluated in the Matrix-String 
approach of \cite{DMVV}) and via a
further S-duality is related to 1-loop 
world-sheet instantons in heterotic theory.
We computed 4 $O(8)$ current correlator in the Yang-Mills theory which receives
contribution only from the 7-brane saddle point as is to be expected from the
fact the $O(8)$ gauge fields live only in the 7-brane world volume, and
showed that
the result is obtainable from $F^4$ 
terms in the 7-brane world volume via AdS/CFT
correspondence. A novel feature we find 
is that also the odd instantons contribute
to such correlators in the Yang-Mills 
side and to $F^4$ term on the string theory
side where it appears as $O(8)$ odd parity quartic invariant. In particular this
shows that BPS instantons  break 
$O(8)$ to $SO(8)$ in contrast with the situation in
10-dimensional type I
theory where $O(32)$ is broken to $SO(32)$ due 
to the non BPS $Z_2$ instanton
\cite{witten4}.

We also discussed some Yang-Mills correlators that do not involve $O(8)$ quantum
numbers. In this case only the even 
instantons contribute and both the saddle points
are relevant. These correlators are given by the $R^4$ couplings (and their
supersymmetric partners) on the 7-brane
world volume and in the 10-dimensional bulk respectively for the two saddle
points:

1) The 7-brane $R^4$ coupling is of the form $t_8 R^4$ and $t_8 (R^2)^2$ that
appears
via T and S-duality at 1-loop in the heterotic theory. The $(-1)$-brane
instanton
contribution to such term is given by 1-loop world sheet instanton effects in
the heterotic theory. These are the 
super invariants $I_1$ and $I_3$ in the notation
of \cite{Tseytlin}, which are related to anomaly cancelling CP odd term and
therefore  are expected to satisfy non-renormalization theorem. 
In the type I' side this result
should be obtainable by computing 8 fermion correlator (these are the zero modes
that are lifted and carry $(-1)$ charge under $O(2)$ R-symmetry) in the
$O(2k)$
matrix integral (where the center of mass has been factored out). It is an
interesting open problem to work out this 
matrix integral and show that it correctly
reproduces the tensor structure of the $I_1$ and $I_3$ invariants as well as the
coefficients in the instanton expansion. 
However we expect this to be true due to 
the non-renrmalization theorem associated with these invariants.

2) More interesting and perhaps somewhat unexpected is the contribution of the
10-dim bulk saddle point. In this case besides the 8 exact supersymmetric and
superconformal zero modes (that transform as $(2,1)_{+1}$ under $SU(2)_A \times
SU(2)_Y 
\times O(2)$) there are 8 more exact fermion zero modes that transform as
$(1,2)_{-1}$. The measure factorizes into 
an integral over the bulk $AdS_5 \times
S^5/Z_2$ together with these 16 fermion zero modes times $SU(k)$ matrix integral
(where $2k$ is the number of instantons) 
appearing in the type IIB context. This is
exactly the situation for the $k$ type IIB $(-1)$ brane instantons and
therefore
should produce $t_8.t_8 R^4$ term (and its supersymmetric partners). This
superinvariant (called $I_0$ in \cite{Tseytlin}) appears at the tree level and
the one loop
torus level in type I and type I' 
(since the computation is the same as in type IIB)
as well as at the tree level in the heterotic theory. 
$I_0$ superinvariant is not
expected to satisfy any non-renormalization theorem. 
Indeed it has been argued in
\cite{Tseytlin} that the heterotic-type I duality demands that $I_0$ receives
correction to all loops already in the 10-dimensional theory. In the type I'
theory under
consideration (i.e. IIB on $T^2/Z_2.(-1)^{F_L}.\Omega$) the $(-1)$-brane
instantons of
the IIB theory are not projected out. Therefore in the large volume limit of 
$T^2$ we expect the $(-1)$-brane instanton 
contributions to the $I_0$ in IIB theory in
10-dim to survive. This is exactly the term which contributes to the
instanton effects in the corresponding
Yang-Mills correlators arising from 
the bulk saddle point. In fact this saddle point
is dominant and goes as $\sqrt{N}$ in the large $N$-limit, due to the fact that
integral
is also over the 2-directions transverse to to 7-brane and its volume goes as
$\sqrt{g_{st}N}$ (the factor $\sqrt{g_{st}}$ 
cancels from the $1/\sqrt{g_{st}}$ that
appears in the leading instanton contribution to $R^4$ \cite{gg, Banks,
Bianchi}).

An interesting open question is what is the fate of the type I' $I_0$
invariant in type I and heterotic theory. By the duality relations one
finds that the leading instanton term in type I' is mapped to
\begin{eqnarray} 
\frac{T_{I'}}{\sqrt{g_{I'}}} e^{-\frac{k}{g_{I'}}} \int d^8
x R_{I'}^4 &\rightarrow & \frac{1}{\sqrt{T_I.g_I}} e^{-
\frac{kT_I}{g_I}}\int d^8 x R_I^4 \nonumber \\ &\rightarrow&
\frac{g_h}{\sqrt{T_h}} e^{-k T_h}\int d^8 x R_h^4 
\label{Ih} 
\end{eqnarray}
where subscripts $I'$, $I$ and $h$ refer to the variables in the three
theories, and $T$ is the volume of the torus. Note that in the heterotic theory
this expression does not make sense due to the odd power of $g_H$. This problem
already appears at the 10-dimensional level, as discussed by \cite{Tseytlin}
with a possible resolution being that both in type I and heterotic side
this term should receive corrections to all orders. 

\subsubsection*{Acknowledgments}

We thank Massimo Bianchi for useful discussions.
This work was supported in part by EC under the TMR
contract ERBFMRX-CT96-0090. M.H.S. would like to thank
Jozef Stefan Institute, SISSA and ICTP for hospitality 
during the course of this work.

\rnc{\Large}{\normalsize}

\end{document}